# Is googling risky? A study on risk perception and experiences of adverse consequences in web search

Helena Häußler, Sebastian Schultheiß & Dirk Lewandowski

Hamburg University of Applied Sciences

Faculty Design, Media and Information

Department of Information

Finkenau 35

22081 Hamburg, Germany

Helena.haeussler@haw-hamburg.de, Sebastian.schultheiss@haw-hamburg.de, Dirk.lewandowski@haw-hamburg.de

## Abstract

Search engines, such as Google, have a considerable impact on society. Therefore, undesirable consequences, such as retrieving incorrect search results, pose a risk to users. Although previous research has reported the adverse outcomes of web search, little is known about how search engine users evaluate those outcomes. In this study, we show which aspects of web search are perceived as risky using a sample ($N = 3,884$) representative of the German Internet population. We found that many participants are often concerned with adverse consequences immediately appearing on the search engine result page. For example, 50.5 % of respondents perceive a risk of not finding what they were looking for, and 45.2 % are concerned about retrieving wrong or misleading information. In contrast, adverse consequences resulting with a delayed impact after the web search are rarely perceived as a risk. Moreover, participants' experiences with adverse consequences are directly related to their risk perception. Our results demonstrate that people perceive risks related to web search. In addition to our study, there is a need for more independent research on the possible detrimental outcomes of web search to monitor and mitigate risks. Apart from risks for individuals, search engines with a massive number of users have an extraordinary impact on society; therefore, the acceptable risks of web search should be discussed.

**Introduction**

When reading newspaper headlines reporting privacy infringements, questionable business practices, and their impacts on search results (e.g., Grind et al., 2019; Knockel & Ruan, 2022; Lomas, 2022), one may wonder about the perpetual popularity of commercial search engines. Researchers in the information retrieval community have constantly been reporting the possible harmful consequences of web search. For example, a person could be exposed to incorrect search results and base relevant decisions on erroneous information, potentially severely threatening their health and well-being (Lin et al., 2020; Maugeri et al., 2022). As undesired outcomes, such as false information, can be regarded as a risk (Aven et al., 2011), the question arises as to whether web search can be considered a risky activity.

The concept of risk has received little attention in the information retrieval field. One research direction associated with information behavior is risk information. A classic example is starting an information-searching process, for example, using a search engine, motivated by hazards such as a flood or pandemic (Shakeri et al., 2018; Wang et al., 2012). However, the literature does not address risks related to *using* search engines. As usage-related risks can have undesirable consequences, there is a need to research the risks of web search. This need is also emphasized by examining large online platforms such as Google and their perspectives on search-related risk. Conducting annual risk assessments will be a future task because search engine operators must comply with the Digital Services Act in the European Union (Cauffman & Goanta, 2021). Most recently, the launch of AI-based conversational search agents by Bing and Google has underscored the urgency of risk in web search since they exacerbate existing risk sources like incorrect results (Enge, 2023). Hence, identifying risks allows researchers to monitor detrimental developments, implement measures to reduce these risks, and hold search engine operators accountable.

Experts and laypeople often come to different conclusions when evaluating risks (e.g., Slovic et al., 2000). Owing to the high number of search engine users with little search expertise (Schultheiß & Lewandowski, 2021b), it is important to capture the risk perceived by laypeople. This study contributes to understanding how the public perceives the risk that various undesirable consequences can happen to them during a web search. To the best of our knowledge, studies on the perceived risks of web search have not yet been published. Moreover, it is unclear how perceived risks are related to personal experiences regarding specific adverse outcomes in web search.

As a first step in this direction, we conducted a study with data gathered via an online questionnaire sent to a representative sample of $N$ = 3,884 German Internet users. In general, descriptive studies of this type allow for describing the distribution of specific properties within a population (Döring et al., 2016, p. 192). In particular, this study provides a picture of risk perceptions and experiences. Owing to its representative nature, the results quantify the degree to which separate problems occur on an individual level. Therefore, they can serve as a basis for discussing and investigating risks on a societal level.

The remainder of this paper is structured as follows. In the next section, we provide an overview of previously identified adverse consequences of web search. This is followed by an outline of the research questions and the methods used in this study. Then, the results are presented and discussed. This paper concludes with a summary and suggestions for future research.

**Literature review**

This section outlines the underlying understanding of risk and perceived risk. Furthermore, we elaborate on the role of risk in web search and describe the adverse outcomes used in the survey.

*Risk and perceived risk*

Risk is a research object in various disciplines, leading to diverging concepts and multiple definitions (Althaus, 2005). In an attempt to review common aspects of risk definitions, Aven et al. (2011) list essential aspects. First, an object must be exposed to a risk source or an event. Such events or human activities may lead to unfavorable and favorable outcomes, although risk is often regarded as a harmful threat to the values at stake. Further, risk describes the possible occurrence of mentioned (adverse) outcomes or events, while the exact circumstances (e.g., actual occurrence, time, severity) remain uncertain. Renn (1998) argues that to speak of risk meaningfully human agency to alter the future is an underlying assumption. Caring about possible future events makes sense if their occurrence can be controlled or modified. Consequently, the probability of the incident events or outcomes serves as a proxy to face the uncertainty. Aven et al. (2011) distinguish frequentist and subjective approaches to determine probability which either build on countable aspects or subjective background knowledge. This distinction gives expression to a fundamental disagreement between risk theories and disciplines.

At one end of the spectrum, realist approaches regard risk as existing in reality and objectively measurable (Bradbury, 1989; Renn, 1992). Predominantly technical disciplines like engineering and epidemiology are dedicated to calculating the probability of undesirable events and the associated damage, according to the over-arching principle of avoiding physical harm (Renn, 1992). Due to the focus on technical probabilities, social factors or processes are excluded from the analysis. This weakness is mitigated in social science disciplines like economics and psychology. Both disciplines agree with realist approaches on the objective existence of risks but consider their social mediation. Based on a rational ideal, economics reviews subjective benefits, and psychology the biased, subjective risk interpretation (Wilkinson, 2006; Zinn, 2008). Therefore, what counts as an undesirable outcome is, to a greater extent, based on personal benefits and interests. Sociological approaches are positioned at the opposite end of the spectrum and can be described as primarily constructivist. Risk is understood predominantly or exclusively as "a product of historically, socially and politically contingent" (Lupton, 1999, p. 35) experiences. A cultural/symbolic approach suggests that a risk notion is shared among the members of a (cultural) group and provides potential for identification (Lupton, 2006). Governmentality approaches and systems theory approaches focus on the discourses, institutions, actors and processes that constitute risk and affect people's behavior and decision-making. Objective risk calculations are addressed but intertwined with risk interpretations generated by social and cultural factors, which extends what is accepted as undesirable consequences (Zinn, 2008). The risk society theory can be situated between the realist and constructivist spectrum. According to their authors, risks are both real and recognized by objective measures and come into existence through social processes and public acknowledgement (Beck, 2016; Giddens, 1999). In particular, risks are seen as a by-product of modernity and technical progress and, in part, fostered by the same scientific community that identifies the probability of risks (Beck, 2016).

Determining the particular risk of an activity or event attempts to control the unknown by applying epistemological knowledge. Realist approaches emphasize measurements carried out by experts who are regarded as holding objective and superior knowledge compared to the public. In contrast, constructivist theories acknowledge the cultural, social and structural influences when determining risks and reject the distinction between experts and laypeople (Zinn, 2008a). Especially psychometric approaches are concerned with differences between the risk assessments by experts and risk perceptions by lay people, predominantly for nuclear and chemical technologies (Slovic, 2000). However, this distinction sparked criticism as the risk judgments by the general public appear inferior to the risk judgments by experts (Bradbury, 1989). Instead, researchers point out that experts are not immune against bias and misjudgments and may come to contrary conclusions, whereas lay people's local and subjective knowledge may provide a sound basis for risk estimation (Giddens, 1999; Lupton, 2006). Nowadays, it may be less a question of which knowledge is superior to the other but rather a question of how this reasoning is shaped and how risk accounts differ on an individual and societal level (Mythen & Walklate, 2006).

*Risk of adverse consequences of web search*

As the risks of negative consequences are particularly grave, we focus on the hazards related to web search. Undesirable outcomes, as identified by the research community, constitute the first approach to determining risks of web search. Recently, Zimmerman et al. (2020) proposed a framework for preventing harm in web search. The authors identified several risks, such as dangerous or lethal health outcomes, destabilized political systems, state-sponsored surveillance, radicalization to extremist communities, and dissemination of dangerous advertisements that can be associated with web search (p. 32). However, due to their focus on precautionary aspects, the authors do not elaborate on the listed risk sources.

First, there are activities whose adverse outcomes become immediately apparent on the SERPs and concern the displayed content, for example, when the search results contain incorrect information. With regard to political information, Metaxa and Torres-Echeverry (2017) found that among a set of search results related to queries about the 2016 U.S. elections, 1.5% led to websites that are known to spread fake news. When looking at politicians' websites alone, the share was 28.9%. In another study, Bondarenko et al. (2021) investigated the first ten organic search results returned by Yandex to popular medical questions. Approximately 44% of the snippets contained incorrect answers. This number can be explained by the preference of the search engine algorithms for positive and confirming search results, which was also observed for Google search results (White, 2013). Therefore, search results may pose a risk of exposing users to biased or one-sided information. Investigating controversial topics, Gezici et al. (2021) did not find bias toward supportive information or counter-arguments. However, they found a slight bias towards positions that agreed with liberal political leaning. Urman et al. (2022) compared search results for candidates of the 2020 U.S. Presidential Primary Elections among six search engines. The authors found that some search engines were more supportive or critical of the candidates. Another result concerns the different preferences of search engines for information types; therefore, users are more likely to be exposed to campaign websites in one search engine and alternative media in another. At the level of website content, Steiner et al. (2020) conducted a content analysis of the first ten organic results to debated subjects collected from five search engines. Across all search engines, the authors found the least diverse information at the first position of the SERP, with diversity increasing strongly until the fifth position.

Personalized search results have also been framed as the cause of bias at the level of confirming personal beliefs. Moreover, the risk consists of potentially missing information in favor of results targeting personal interests and information needs; these concerns are often subsumed under the term "filter bubble." However, the lack of search results owing to personalization is difficult to observe. Lai and Luczak-Roesch (2019) compared the relevance ratings of personalized and non-personalized search results. They found that up to 20% of the search results absent in the personalized SERP were rated as relevant by the participants. However, the overall effect of personalization on search results is marginal (Puschmann, 2019; Robertson et al., 2018; Trielli & Diakopoulos, 2022). Recently, an analysis of non-personalized search results suggested that a certain degree of randomization affects the retrieved search results and their ranking (Urman et al., 2022).

The influence exerted by search engine operators on search results may have an immediate effect on the search results. Notably, Google's violations of competition and antitrust laws have been documented in numerous sentences and penalties. The accusations concern, for example, the abuse of its market dominance as a search engine for the preferred placement of its services (e.g., Google Shopping) to the detriment of competitors (Lewandowski & Sünkler, 2013). In addition, according to a recent analysis, Google covers up to 41% of the SERP with its own content, including "people also ask," services, such as a flight planner, or knowledge panels, resulting in decreased visibility of organic results (Jeffries & Yin, 2020).

Other undesirable consequences at the SERP level could potentially affect the ranking of the results. Typically, search engines display results according to some concept of relevance, although relevance may be interpreted differently through the lens of society, individual users, and search engine operators (Sundin et al., 2021). Studies have found considerable discrepancies between search results ranked by search engines and those favored by users (Bar-Ilan et al., 2007; Zhitomirsky-Geffet et al., 2016). Moreover, users may interpret relevance differently at different times (Zhitomirsky-Geffet et al., 2016). Consequently, it may be complicated for users to recognize relevant search results. The number of presented search results may also add to this. Chiravirakul and Payne (2014) found that time pressure aggravates the difficulty of selecting search results from a large result set. As Foulds et al. (2021) show, recognizing relevant search results is more difficult for users if advertisements are present because they increase the number of items on a SERP. Advertisements on the SERP also have the potential to lead users to websites they do not want to access because many users lack the ability to reliably distinguish sponsored from organic results (Lewandowski et al., 2018; Schultheiß & Lewandowski, 2021a).

Furthermore, in the context of web search it is relevant to look at the undesirable consequences that may result from an event's or action's negative that actually occurred. Such delayed impacts can be found in studies concerned with making decisions or forming beliefs. Medical and health information is widely used because it emphasizes that something valuable—health or life—is at stake. Generally, laypeople have difficulties estimating the efficacy of medical treatments, with a tendency to overrate it (Hashavit et al., 2021). When exposed to search results containing incorrect information about medical treatments, many people reject treatments that are in fact helpful (Pogacar et al., 2017). Similar results were obtained for the information displayed as featured snippets at the top of the SERP. Incorrect statements in featured snippets led to a decrease in the answer accuracy of approximately 30% (Bink et al., 2022). Fischer et al. (2021) found that weak code snippets displayed at the top of a SERP are often incorporated into other insecure software codes. Thus, low-quality search results can lead to insecure software development. Search results may also affect people's opinion-making. Allam et al. (2014) showed that being predominantly exposed to antivaccination information in search results led to decreased knowledge about vaccinations and increased fear of side effects. However, participants who accessed high-quality and neutral information about abortion were less concerned about the associated side effects (Forbes et al., 2021). Search results may seem correct at first and turn out to be incorrect later, such as fake e-commerce shops accessed via search results (Carpineto & Romano, 2017; Leontiadis et al., 2014).

Additionally, some adverse outcomes with immediate and delayed impacts may be difficult to relate as a cause to a singular activity or event. Privacy infringements are often unknown to those exposed and thus pose a risk associated with web search (Zimmerman, Thorpe, et al., 2020). Massive online behavior tracking, as pushed excessively by Google (Englehardt & Narayanan, 2016), violates the right to privacy. Recently, Google was fined by 40 U.S. states for tracking the location of users who explicitly disagreed with location sharing (Collins & Gordon, 2022).

Overall, these studies support the notion that several risks are associated with web search and that many of them can be described in objective terms. However, those risks are also socially mediated, e.g., through the public debate on misinformation or national elections. Therefore, risks in web search are best conceptualized in the midst of the spectrum of realist and constructivist approaches. Furthermore, it is an interesting question to what extent people are aware of those risks at the individual level, apart from the societal impact. The risks of web search identified in the mentioned studies form the basis for the survey construction.

**Research questions**

This descriptive study aimed to determine the perceived risk of using a search engine and examine its possible causes.

Therefore, we asked the following research questions:

RQ1: Which adverse consequences of web search do search engine users perceive?

RQ2: How does a user's personal experience with a particular adverse consequence increase its perceived risk?

To identify group differences within the data, we asked the following question:

RQ3: Which factors influence a user's perceived risk of adverse consequences of web search?

**Methods**

Following a realist and weak constructivist risk concept, we aimed at adding the perspective of the general public to the risks of web search identified by domain experts. We designed a questionnaire consisting of two consecutive questions. A pre-test with $N$ = 10 participants was conducted to test for comprehension. Afterwards, the questionnaire was distributed via e-mail to the users in a registered panel and via randomized layers to the visitors of 20 popular websites from the sectors of finance, news and entertainment. In total, a sample ($N$ = 3,884) representative of the German Internet population ($N$ = 61,150,000) took part in the study. The market research company Fittkau & Maaß Consulting conducted the survey in April and May of 2022. All files from the study are available from the OSF repository (https://dx.doi.org/10.17605/OSF.IO/7AF2E).

Based on the risks of web search reported in the literature review, we compiled a list of 13 possible web search consequences (Table 1). The list items were selected to represent a broad range of potential negative consequences that individuals may encounter at the SERP level. They are varying in their degree of probability, severity and immediacy. Thereby, consequences were included that received attention in media and politics (e.g., wrong and biased search results), as well as less prominent ones (e.g., missing and contradictive results). The item "That I don't find what I was looking for" was included following previous studies (Dutton et al., 2017), and the aspect "That search results are not useful" targeted the mission of Google to provide "helpful" results.[1]

**Table 1**

*List of adverse consequences of web search used in the study*

| Question I: "Suppose you are using a search engine (e.g., Google, Bing) for a specific search: what could happen to you? Check all that apply." | Question II: "Thinking of your personal experiences with search engines (e.g., Google, Bing), which of the following [items] has already happened to you during usage/search? Please specify: often, 1 or 2 times, never." |
|---|---|
| That I don't find what I was looking for | That I didn't find what I was looking for |
| That I receive wrong or misleading information | That I received wrong or misleading information |
| That the search engine operators' (e.g., Google, Bing) interests influence search results | That the search engine operators' (e.g., Google, Bing) interests influenced search results |

---

[1] https://www.google.com/search/howsearchworks/our-approach/.

| | |
|---|---|
| That the best results are at the bottom of the result list | That the best results were at the bottom of the result list |
| That I receive too many search results | That I received too many search results |
| That search results are not useful | That search results were not useful |
| That important search results are missing | That important search results were missing |
| That search results are biased, represent only one perspective | That search results were biased, represented only one perspective |
| That I can't distinguish sponsored search results from non-sponsored results and accidentally click on them | That I couldn't distinguish sponsored search results from non-sponsored results and accidentally clicked on them |
| That search results are contradictive, and I don't know what's correct | That search results were contradictive, and I didn't know what was correct |
| That I don't recognize the relevance of a search result | That I didn't recognize the relevance of a search result |
| That search results later turn out to be wrong | That search results later turned out to be wrong |
| That I make a bad decision based on the search results | That I made a bad decision based on the search results |
| Something else, please specify [text answer] | |
| Nothing | |

The first question aimed to collect perceived risk by asking users to choose the listed items that they deemed a risk. The exact wording was: "Suppose you are using a search engine (e.g., Google, Bing) for a specific search: what could happen to you?" [translated from German]. Participants were asked to check all the items that apply in their view. Participants should not be sensitized to perceive risk through the wording, thus the term "risk" was not used in the question (Kohring, 2004, p. 209). To account for a subjective risk assessment, we provided an additional free text option to allow naming of the risks that were not included in the list. Finally, the option "Nothing" was added to allow the participants to indicate that no risks were perceived.

Since personal experience may be an important indicator for risk perception, the occurrence and frequency of personal experiences with the items given in Question 1 was surveyed in Question 2. The wording of the question was: "Thinking of your personal experiences with search engines (e.g., Google, Bing), which of the following [items] has already happened to you during usage/search?" [translated from German]. For each item, respondents were asked to specify if it had happened to them either never, rarely (one or two times), or often (three or more times).

We performed different statistical procedures to analyze the data. Chi-square tests of association were performed to examine the relationships between perceived risk and personal experience, gender and perceived risk, and age group and perceived risk. To analyze the results regarding age differences, we divided the participants into age groups according to Western generational groupings (Burclaff, 2021). The youngest is Generation Z (born between 1997 and 2012), followed by Generation Y (born between 1981 and 1996), Generation X (born between 1965 and 1980), the Baby Boomer Generation (born between 1946 and 1964), and the Silent Generation (born between 1928 and 1945). The alpha level for all the statistical tests was set at .05. Data analysis was performed using IBM SPSS

Statistics 27. The answers added via the free text option were coded into subject categories by a researcher and a student assistant. Following a consensus coding approach, divergences were identified, discussed and agreed upon.

## Results

### *Description of the sample*

The sample consisted of $N = 3,884$ Internet users from Germany. Of them, 47.8% were women ($N = 1,856$) and 52.2% men ($N = 2,028$). The respondents were aged between 18 and 92 years, with a mean age of 45 years ($SD = 15.6$).

### *Perceived risk*

We asked respondents what adverse consequences they perceived during a web search (Question 1) and what consequences they had already experienced (Question 2). As shown in Table 2, most users (50.5%) perceived a risk of not finding what they were looking for. In contrast, the least frequently selected item concerned making bad decisions based on the search results (9.2%). Of all the participants, 9.1% stated that they did not perceive risks at all.

**Table 2**

*Perceived and experienced risks*

| Adverse consequences | Question I (perceived risk) % | Question II (personal experience) | | |
|---|---|---|---|---|
| | | Often (3 or more times) % | Rarely (1 or 2 times) % | Never % |
| That I don't find what I was looking for | 50.5 | 45.3 | 41.5 | 10.7 |
| That I receive wrong or misleading information | 45.2 | 41.5 | 37.6 | 16.0 |
| That the search engine operators' (e.g., Google, Bing) interests influence search results | 43.1 | 41.5 | 32.0 | 20.8 |
| That the best results are at the bottom of the result list | 37.4 | 38.2 | 40.4 | 17.6 |
| That I receive too many search results | 36.5 | 55.1 | 20.8 | 19.1 |
| That search results are not useful | 34.4 | 38.1 | 40.7 | 16.6 |
| That important search results are missing | 33.0 | 30.8 | 41.2 | 23.0 |
| That search results are biased, represent only one perspective | 27.2 | 31.0 | 38.2 | 24.5 |
| That I can't distinguish sponsored search results from non-sponsored results and accidentally clicked on them | 24.4 | 23.3 | 36.3 | 34.8 |
| That search results are contradictive and I don't know what's correct | 21.5 | 26.2 | 39.6 | 28.4 |
| That I don't recognize the relevance of a search result | 17.7 | 16.8 | 40.1 | 37.8 |
| That search results later turn out to be wrong | 14.1 | 12.9 | 42.1 | 38.6 |
| That I make a bad decision based on the search results | 9.2 | 6.3 | 31.0 | 56.4 |
| Something else, please specify [text answer] | 1.8 | - | - | - |
| Nothing | 8.9 | - | - | - |

*Note.* The highest value of each response option is given in bold.

Participants could add more adverse outcomes of the web search via the free text option, and 1.8% ($N = 71$) of them did. In total, $N = 117$ statements were made, and some participants made several statements. We excluded $N = 13$ statements because they were incomprehensible. Most open answers focused on immediate adverse consequences, such as being exposed to advertising in search results ($N = 12$), misinterpretation of the search intent leading to irrelevant search results ($N = 9$), being redirected to spam websites ($N = 7$), receiving outdated or less recent search results ($N = 7$), and obtaining unavailable or nonexistent search results ($N = 6$). Moreover, the respondents emphasized suspected general manipulation by the search engine operator or unspecified third parties ($N = 9$). The risks of privacy infringement and tracking were also mentioned ($N = 17$). Although the list provided in Question 1 included only negative consequences, the question formulation also allowed for positive comments. Thus, people stated that they would find (unexpectedly) what they were looking for ($N = 15$).

As respondents could select as many adverse consequences as apply in their view, most participants selected multiple outcomes. On average, the participants selected four items ($SD = 3.02$). As shown in Table 3, the higher the number of selected consequences, the lower the number of participants. Thus, few participants selected ten or more consequences (6% in total).

**Table 3**

*Frequency table of selected adverse consequences*

| Number of selected adverse consequences | N | % | % (cumulative) |
|---|---|---|---|
| 0 | 371 | 9.5 | 9.5 |
| 1 | 580 | 14.9 | 24.5 |
| 2 | 507 | 13.1 | 37.5 |
| 3 | 573 | 14.7 | 52.3 |
| 4 | 409 | 10.5 | 62.8 |
| 5 | 404 | 10.4 | 73.2 |
| 6 | 320 | 8.2 | 81.4 |
| 7 | 226 | 5.8 | 87.3 |
| 8 | 163 | 4.2 | 91.5 |
| 9 | 99 | 2.5 | 94.0 |
| 10 | 71 | 1.8 | 95.8 |
| 11 | 58 | 1.5 | 97.3 |
| 12 | 53 | 1.4 | 98.7 |
| 13 | 51 | 1.3 | 100 |

*Personal experiences with adverse consequences*

The second question asked about the participants' personal experiences with adverse consequences listed in Question 1. The participants stated how frequently the individual incidents happened to them.

From the columns for Question 2 in Table 2, more than half of the sample (55.1%) has often received too many search results, a number that contrasts with the 36.5% who perceive it as a risk in web search. A small number of 6.3% of the participants often made bad decisions based on search results. Noticing that search results later turned out to be incorrect was reported by 55.0% ("often" and "1 or 2 times" together), which is a high value compared to 14.1% of participants who perceived this a risk. Looking at the response option "never," 56.4% of the participants stated that they never made a bad decision based on search results, and 38.6% reported having never encountered that search results later turned out to be incorrect.

*Relationship between perceived risk and personal experience*

Personal experiences with the individual adverse consequences of web search are likely to affect risk perception. To test for the relationship between perceived risk (Question 1) and personal experience (Question 2), chi-square tests of association (independent samples) were conducted. For Question 2, the response options "1 or 2 times" and "often" were summarized as "experience." This association was significant for each adverse consequence (Table 4). The effect sizes were small to medium, according to Cohen (1992). Among those who reported that they had already experienced not finding what they were looking for, 55.8% perceived the corresponding risk. Among those who had not yet had the same experience, only 13.7% perceived the risk.

**Table 4**

*Chi-square tests of association between perceived risk and personal experience*

| Adverse consequences | Subjects who perceive a risk | | Results of chi-square tests |
|---|---|---|---|
| | when experience has been made (%) | when experience has not been made (%) | |
| That I don't find what I was looking for | 55.8 | 13.7 | ($\chi^2$(1, $N$ = 3,788) = 262.068, $p$ < .001), small effect size ($\varphi$ = .263) |
| That the search engine operators' (e.g., Google, Bing) interests influence search results | 52.0 | 15.8 | ($\chi^2$(1, $N$ = 3,661) = 334.688, $p$ < .001), medium effect size ($\varphi$ = .302) |
| That I receive wrong or misleading information | 51.4 | 19.9 | ($\chi^2$(1, $N$ = 3,694) = 206.991, $p$ < .001), small effect size ($\varphi$ = .237) |
| That the best results are at the bottom of the result list | 45.1 | 7.7 | ($\chi^2$(1, $N$ = 3,696) = 354.405, $p$ < .001), medium effect size ($\varphi$ = .310) |
| That I receive too many search results | 44.9 | 10.0 | ($\chi^2$(1, $N$ = 3,686) = 287.589, $p$ < .001), small effect size ($\varphi$ = .278) |
| That search results are not useful | 41.0 | 11.1 | ($\chi^2$(1, $N$ = 3,689) = 270.642, $p$ < .001), small effect size ($\varphi$ = .271) |
| That important search results are missing | 39.8 | 14.5 | ($\chi^2$(1, $N$ = 3,702) = 149.654, $p$ < .001), small effect size ($\varphi$ = .201) |
| That search results are biased, represent only one perspective | 36.8 | 5.5 | ($\chi^2$(1, $N$ = 3,669) = 443.461, $p$ < .001), medium effect size ($\varphi$ = .348) |
| That I can't distinguish sponsored search results from non-sponsored results and accidentally click on them | 34.7 | 8.9 | ($\chi^2$(1, $N$ = 3,683) = 243.502, $p$ < .001), small effect size ($\varphi$ = .259) |
| That search results are contradictive and I don't know what's correct | 28.7 | 7.2 | ($\chi^2$(1, $N$ = 3,655) = 206.719, $p$ < .001), small effect size ($\varphi$ = .238) |
| That I don't recognize the relevance of a search result | 26.5 | 6.0 | ($\chi^2$(1, $N$ = 3,677) = 247.935, $p$ < .001), small effect size ($\varphi$ = .260) |
| That search results later turn out to be wrong | 20.8 | 5.9 | ($\chi^2$(1, $N$ = 3,635) = 157.179, $p$ < .001), small effect size ($\varphi$ = .208) |
| That I make a bad decision based on the search results | 17.4 | 4.2 | ($\chi^2$(1, $N$ = 3,638) = 175.445, $p$ < .001), small effect size ($\varphi$ = .220) |

*Factors influencing perceived risk*

To analyze whether certain groups were more prone to perceive the risk of adverse consequences of web search, we conducted chi-square tests of association again. We tested whether there is a relationship between gender and perceived risk and between age groups and perceived risk (RQ3).

For 7 of the 13 listed adverse outcomes, there was a significant relationship between perceived risk and gender, as shown in Table 5. For the five adverse consequences, male participants more often indicated that they may occur during a web search. At the same time, women were more likely to perceive the risk of overload and insecurity when faced with contradictory results. The effect sizes were small, ranging between .033 and .081.

**Table 5**

*Chi-square tests of association between perceived risk and gender*

| Adverse consequences | Question I (perceived risk) | | Results of chi-square tests |
|---|---|---|---|
| | **Female (%)** | **Male (%)** | |
| That I don't find what I was looking for | 50.0 | 51.0 | ($\chi^2$(1, $N$ = 3,883) = .398, $p$ = .528) |
| That I receive wrong or misleading information | 43.5 | 46.8 | ($\chi^2$(1, $N$ = 3,884) = 4.158, $p$ = .041), small effect size ($\varphi$ = .033) |
| That the search engine operators' (e.g., Google, Bing) interests influence search results | 38.9 | 46.9 | ($\chi^2$(1, $N$ = 3,885) = 25.109, $p$ < .001), small effect size ($\varphi$ = .080) |
| That I receive too many search results | 38.4 | 34.8 | ($\chi^2$(1, $N$ = 3,883) = 5.478, $p$ = .019), small effect size ($\varphi$ = .038) |
| That search results are not useful | 34.1 | 34.8 | ($\chi^2$(1, $N$ = 3,883) = .206, $p$ = .650) |
| That the best results are at the bottom of the result list | 33.4 | 41.2 | ($\chi^2$(1, $N$ = 3,885) = 25.195, $p$ < .001), small effect size ($\varphi$ = .081) |
| That important search results are missing | 31.6 | 34.2 | ($\chi^2$(1, $N$ = 3,884) = 2.950, $p$ = .086) |
| That search results are biased, represent only one perspective | 26.2 | 28.1 | ($\chi^2$(1, $N$ = 3,884) = 1.618, $p$ = .203) |
| That search results are contradictive and I don't know what's correct | 23.5 | 19.6 | ($\chi^2$(1, $N$ = 3,883) = 8.872, $p$ < .01), small effect size ($\varphi$ = .048) |
| That I can't distinguish sponsored search results from non-sponsored results and accidentally click on them | 22.3 | 26.5 | ($\chi^2$(1, $N$ = 3,885) = 9.318, $p$ < .01), small effect size ($\varphi$ = .049) |
| That I don't recognize the relevance of a search result | 16.4 | 18.8 | ($\chi^2$(1, $N$ = 3,884) = 3.844, $p$ = .05) |
| That search results later turn out to be wrong | 13.8 | 14.4 | ($\chi^2$(1, $N$ = 3,884) = .279, $p$ = .597) |
| That I make a bad decision based on the search results | 7.9 | 10.3 | ($\chi^2$(1, $N$ = 3,884) = 6.883, $p$ < .001), small effect size ($\varphi$ = .042) |

*Note.* The highest value of each *significant* difference is given in bold.

With regard to age groups, for most (8 out of 13) of the adverse outcomes, there was a significant relationship between perceived risk and age (**Fehler! Verweisquelle konnte nicht gefunden werden.**). The effect sizes were small, ranging between .067 and .157. Among generations, differences in perceived risks became apparent. Younger participants from Generation Z seemed to struggle with contradictory search results (27.8%) and recognizing the relevance of a search result (23.5%). For five adverse consequences, such as influence by the search engine operator, we found that Generation X members had higher percentages than the other age groups. Receiving not useful search results (37.8%), is most often perceived as a risk by respondents from the Generation of Baby Boomers.

**Table 6**

*Chi-square tests of association between perceived risk and age groups*

| Adverse consequences | Question I (perceived risk) | | | | | Results of chi-square tests |
|---|---|---|---|---|---|---|
| | Generation Z (%) | Generation Y (%) | Generation X (%) | Generation Boomer (%) | Generation Silent (%) | |
| That I don't find what I was looking for | 50.8 | 47.5 | 51.4 | 52.1 | 59.8 | ($\chi^2$(4, $N$ = 3,884) = 8.158, $p$ = .086) |
| That the search engine operators' (e.g., Google, Bing) interests influence search results | 27.7 | 39.0 | 51.1 | 48.0 | 36.1 | ($\chi^2$(4, $N$ = 3,885) = 95.548, $p$ < .001), small effect size (Cramer's $V$ = .157) |
| That I receive wrong or misleading information | 44.0 | 42.1 | 49.4 | 44.6 | 33.7 | ($\chi^2$(4, $N$ = 3,885) = 18.980, $p$ < .001), small effect size (Cramer's $V$ = .070) |
| That the best results are at the bottom of the result list | 30.2 | 39.4 | 41.0 | 34.8 | 24.2 | ($\chi^2$(4, $N$ = 3,883) = 29.588, $p$ < .001), small effect size (Cramer's $V$ = .087) |
| That I receive too many search results | 33.4 | 35.6 | 36.8 | 39.3 | 38.6 | ($\chi^2$(4, $N$ = 3,884) = 5.354, $p$ = .253) |
| That search results are not useful | 28.4 | 32.1 | 37.0 | 37.8 | 33.7 | ($\chi^2$(4, $N$ = 3,885) = 19.125, $p$ < .001), small effect size (Cramer's $V$ = .070) |
| That important search results are missing | 34.2 | 30.6 | 34.9 | 33.2 | 24.1 | ($\chi^2$(4, $N$ = 3,884) = 8.612, $p$ = .072) |
| That search results are biased, represent only one perspective | 25.2 | 25.0 | 31.2 | 25.4 | 21.7 | ($\chi^2$(4, $N$ = 3,885) = 17.257, $p$ = .002), small effect size (Cramer's $V$ = .067) |
| That I can't distinguish sponsored search results from non-sponsored results and accidentally click on them | 15.7 | 25.7 | 27.5 | 23.5 | 21.7 | ($\chi^2$(4, $N$ = 3,885) = 30.214, $p$ < .001), small effect size (Cramer's $V$ = .088) |
| That search results are contradictive and I don't know what's correct | 27.8 | 23.6 | 19.3 | 17.6 | 22.9 | ($\chi^2$(4, $N$ = 3,885) = 26.564, $p$ < .001), small effect size (Cramer's $V$ = .083) |
| That I don't recognize the relevance of a search result | 23.5 | 19.0 | 17.0 | 13.8 | 10.8 | ($\chi^2$(4, $N$ = 3,886) = 24.735, $p$ < .001), small effect size (Cramer's $V$ = .080) |
| That search results later turn out to be wrong | 13.4 | 14.3 | 13.9 | 14.8 | 12.2 | ($\chi^2$(4, $N$ = 3,884) = .823, $p$ = .935) |

| That I make a bad decision based on the search results | 9.8 | 10.7 | 8.4 | 8.2 | 3.7 | ($\chi^2$(4, N = 3,884) = 6.61, p = .158) |

*Note.* The highest value of each *significant* difference is given in bold.

## Discussion

When asked about the adverse consequences of web search, respondents regarded many of them as possible (RQ1). Consequences with immediate effects, such as not finding the desired information or retrieving false results, are more often perceived as risks than outcomes with delayed effects, such as subsequent decisions. Nevertheless, approximately ten percent of the respondents did not perceive any risk associated with the web search. Personal experience emphasizes the tendency toward perceived risk since more participants reported having often experienced adverse outcomes with those immediate effects. Indeed, a significant positive correlation was found between the two questions, indicating that the more often one has experienced a specific adverse outcome, the more likely one perceives it as a risk (RQ2). Additionally, there was a significant correlation between many listed adverse outcomes and gender and age groups (RQ3).

The most rated adverse consequence was not finding what one was looking for as most participants had already experienced this consequence in the past. Previous studies have arrived at a similar conclusion that the majority of the surveyed people have experienced a deviation from the objective of finding the desired information (Dutton et al., 2017; Schultheiß & Lewandowski, 2021c). However, moments of serendipity and unexpectedly finding the desired information were mentioned in the open answers and are common to the experiences of search engine users as well (Dutton et al., 2017). Further, many respondents consider the influence of search engine operators on search results to be a considerable risk, as was supplemented by open answers. Surveys indicate that users expect search results to serve their interests and critically view the possibility of adapting the ranking to the operators' interests (European Commission, 2016).

The perceived risk of coming across incorrect or misleading search results is backed by personal experience, which is contrary to the findings from the Digital News Report 2018. Respondents from the global sample were much more concerned about specific situations of misinformation online, although they hardly were aware of precise incidents of being exposed to misinformation (Newman et al., 2018). Similarly, a survey among young Germans about general Internet risks found the perceived risk for adverse consequences to be considerably higher than practical experience (Deutsches Institut für Vertrauen und Sicherheit im Internet & SINUS-Institut Heidelberg, 2018). However, adverse consequences with immediate impacts are often perceived as a risk, which was also found in the present study.

Interestingly, the demographics of our study revealed that women less often perceived the risk of adverse consequences in web search. This finding is contrary to numerous previous studies that sketch an image of cautious women recognizing many risks and adventurous men perceiving fewer risks (e.g., Chauvin et al., 2007; Finucane et al., 2000). It also seems noteworthy that men more often perceived the risk of retrieving false information, since they appeared confident about the accuracy of web search results (Taylor & Dalal, 2017). Additionally, age may contribute to perceived risk, as older persons tend to perceive more risks (Chauvin et al., 2007), which was not observed in the present study.

The differences between the perceived risk of adverse outcomes and personal experiences may be explained by the wording of the first question, which neutrally asked what could happen. In contrast, the other surveys explicitly emphasized the negative aspects due to speaking of "concern" and "risk" or "fear" (Deutsches Institut für Vertrauen und Sicherheit im Internet & SINUS-Institut Heidelberg, 2018; Newman et al., 2018). These words may yield the expectation of endorsing the supposed norm, although individuals struggle to apply their own epistemological knowledge to such a hazard.

Being able to report encounters with incorrect information implies recognition of errors. According to the Eurobarometer survey (2018), more than half of the surveyed European Internet and online platform users felt confident in identifying incorrect information online. However, such surveys rely on self-reported data and do not offer an opportunity to verify actual experiences. Clicking inadvertently on sponsored results, receiving biased results, or learning later that search results are incorrect also requires specific attention to those occurrences. Other consequences may appear without notice, such as the absence of search results or the influence of the search engine operator. Contrary to previous surveys, the listed adverse outcomes of web search were not developed with a qualitative sample before the questionnaire. Therefore, not all items are clearly discriminated; that is, not finding what one was looking for may also touch upon the ranking and result overload. Another limitation is the answer options of Question 2, which should be more consistent and allow for smaller ranges between "often" and "1 or 2 times."

Additional qualitative studies should be conducted in the future to better distinguish adverse consequences of web search. On a larger scale, search engine research should provide evidence for estimating the possibilities and probabilities of specific adverse consequences, both by experts and lay people. Cases of severe harm associated with search engine use should be quantified, that is, counted and documented. In addition, future surveys should also include indicators about individuals' values and social class, as well as acceptable and desirable risks of web search. Typically, acceptable risks are surveyed based on the benefits of an activity or event and contrasted against each other (Slovic, 2000). It is likely that search engine users concede the risk of specific undesirable outcomes when faced with convenient access to online information.

The present study may increase attention to risk issues among search engine researchers. Risk perceptions are subject to dynamic circumstances caused by new scientific evidence, personal experiences or media coverage (Lupton, 2006). Consequently, it is necessary to conduct large-scale comparative studies that monitor those developments among experts and the public. In addition, research on risk perceptions should incorporate both the individual and societal levels. People may come to different conclusions when evaluating the risk of adverse outcomes for themselves and society (Wilkinson, 2006). Recently, the Digital Services Act (DSA), approved by the European Parliament and Council of Europe, draws attention to risks with impacts on society. Search engines with more than 45 million monthly users, such as Google and Bing, are required to perform an annual assessment of systemic risks, that is, the dissemination of illegal content, threatening fundamental rights, democratic processes and public health (Digital Service Act, 2022). However, as Cauffman and Goanta (2021) stated, the responsibility for identifying (and mitigating) risks is left to the platforms (p. 770). The example of Meta (Facebook at that time) provides an impression of how large platforms handle risk assessments. In September 2021, leaked internal research revealed a range of harms on Facebook and Instagram. The analysis showed that Instagram content negatively affects the body image of teenage girls and gives rise to eating disorders and depression (Wells et al., 2021). Consequently, there is a need for independent research, which will be facilitated by access to platform data for research purposes, also within the DSA. Furthermore, search engines with smaller numbers of users, such as DuckDuckGo, should not be overlooked because they are not immune to the same risks (Makhortykh et al., 2020).

The risks perceived by the public may indicate existing consciousness and guide interventions accordingly. Careful risk communication can raise awareness for overlooked negative consequences like the ones with delayed effects. As the small number of open answers about the risk of tracking and privacy infringement indicates, public attention may not be sufficient if individuals do not relate it to web search. Since it may be difficult for individuals to keep track of outcomes, the connection between search results and subsequent decisions or events should be made visible. Furthermore, adequate user information can illustrate the interference between risks at the individual and societal level, such as destabilized political systems and radicalization processes.

## Conclusion

Taking motivation from the recent press and research reports of undesirable consequences related to search engines, this study raised the question of how to assess the risk of web search. The survey respondents indicated the perceived risk of several adverse consequences and reported how often a separate incident happened to them. The results show that most participants were concerned with not finding the desired information and being exposed to the influence of search engine operators. On average, people selected 4 out of 13 adverse consequences, suggesting a moderate level of risk. People who have experienced a specific adverse outcome are significantly more likely to perceive a risk that the same may happen to them again. We also found a significant relationship between perceived risk and gender and age groups for many adverse outcomes.

The results of this study may inspire subsequent research in the information retrieval and information science communities. Approaches are needed to help determine the possible adverse outcomes of web search and to clarify their dimensions for individuals and society. Then, based on objective probabilities and public risk perceptions, the insights can be used to assess which risks should be reduced and which are acceptable. In particular, search engines that reach a high number of users should be held accountable for the risks, but this should be complemented by independent research.

## Acknowledgements

The authors would like to thank Fittkau & Maaß Consulting for conducting the online survey and for giving valuable feedback on the research design.

## Research Data

All files from the online survey are available from the OSF repository (https://dx.doi.org/10.17605/OSF.IO/7AF2E).